\documentclass[%
reprint,
amsmath,amssymb,
aps,onecolumn,eqsecnum]{article}

\usepackage{float}
\usepackage{graphicx}
\usepackage{dcolumn}
\usepackage{bm}
\usepackage{dirtytalk}
\usepackage{listings}
\usepackage{tikz-cd}
\usepackage{amssymb}
\usepackage{amsmath}
\usepackage{url}
\usepackage{amsfonts}
\usepackage{caption}
\numberwithin{equation}{section}
\usepackage{amsthm}

\usepackage{graphicx}
\usepackage{amssymb}
\usepackage{url}
\usepackage[utf8]{inputenc}
\usepackage[english]{babel}
\usepackage[left=2.5cm,right=2.5cm,top=2cm,bottom=2cm]{geometry}
\usepackage{fancyhdr}
\usepackage{enumerate}
\usepackage{amssymb}
\usepackage{amsfonts}
\usepackage{amsmath}
\usepackage{perpage}
\MakePerPage{footnote}
\usepackage[titletoc,title]{appendix}

\usepackage{multicol}
\usepackage{enumitem}
\setlist[enumerate,1]{start=0} 
\lstdefinestyle{myCustomPythonStyle}{
	language=Python,
	stepnumber=1,
	numbers=left,
	numbersep=10pt,
	tabsize=4,
	showspaces=false,
	showstringspaces=false
}
\title{
	\textsc{Quantum Trajectories in Entropic Dynamics}
}

\author{
	Nicholas Carrara\\State University of New York at Albany. 
	\and 
}

\date{
	\today
}

\begin{document}
	
	\maketitle
	
	\hrule



\abstract{Entropic Dynamics \cite{Caticha_new} is a framework for deriving the laws of physics from entropic inference.  In an (ED) of particles, the central assumption is that particles have definite yet unknown positions.  By appealing to certain symmetries, one can derive a quantum mechanics of scalar particles and particles with spin, in which the trajectories of the particles are given by a stochastic equation.  This is much like Nelson's stochastic mechanics \cite{Nelson} which also assumes a fluctuating particle as the basis of the microstates.  The uniqueness of ED as an entropic inference of particles allows one to continuously transition between fluctuating particles and the smooth trajectories assumed in Bohmian mechanics \cite{Bohm}.  In this work we explore the consequences of the ED framework by studying the trajectories of particles in the continuum between stochastic and Bohmian limits in the context of a few physical examples, which include the double slit and Stern-Gerlach experiments.
}











\section{Introduction}
Entropic Dynamics (ED) is a unique approach to foundational quantum mechanics with its emphasis on entropic inference.  It is argued, simply, that physics cannot be an exception to the rules of inductive reasoning;  physics is constrained to be consistent with the rules for inference.  (ED) is an exercise in deriving physical laws from inductive inference.  The main assumption in (ED) is that particles have definite yet unknown positions, and that these positions determine entirely the \textit{ontic} elements of the theory.  All other \textit{observable} quantities, such as momentum, spin, electric charge, etc., are necessarily epistemic.  This is a slight departure from the Copenhagen interpretation, which claims that particles have \textit{no} properties until they are measured.  Other foundational approaches, such as the Bohmian \cite{Bohm} (or casual interpretation) and Nelson's stochastic mechanics \cite{Nelson}, also assume ontic positions for particles.  These approaches however also give onticity to the macroscopic variables, such as the wave function $\psi(x)$, and the probability distribution $\rho(x) = |\psi(x)|^2$.  In (ED) the macroscopic variables are also necessarily epistemic.\\
\indent  In the causal approach particles are assumed to follow smooth trajectories whose velocities are determined by the probability flow \cite{DHK_Spin}.  In this way it is a deterministic theory with respect to particle positions; given initial conditions the trajectory of the particle is known exactly. The uncertainty in positions can therefore only be blamed on not knowing the initial conditions, or not knowing the proper Hamiltonian.  While entirely consistent with quantum mechanics, it is impossible to determine whether the Bohmian interpretation is general enough with respect to particle trajectories, since we cannot set up experiments in which the Hamiltonian is known exactly.  Therefore, fluctuations can always be blamed on a lack of this information and not, necessarily, on some sub-quantum effect that (BM) has failed to include.  Nelson's stochastic mechanics (NSM) is more general in this regard, since it begins with a stochastic equation for the motion of particles and proceeds to derive the dynamics of the macroscopic variables from these assumptions.  While in this way (NSM) is more general than (BM), it singles out a particular sub-quantum dynamics for particles which is a Brownian motion.  Much like the casual picture, (NSM) gives ontic privilege to the macroscopic variables which is part of the reason for its downfall \cite{Wallstrom}.  While (NSM) can obtain the Bohmian limit, simply by sending the fluctuations to zero, they cannot necessarily motivate the generalized dynamics offered by (ED).\\
\indent Entropic Dynamics allows for a more generalized sub-quantum dynamics which includes the (NSM) and (BM) limits as special cases.  Particle trajectories are derived from the principle of maximum entropy by incorporating uncertainty in their motion for small steps $\Delta t$.  Once we specify the relevant constraints in the problem, we can find the transition probability for these small steps $P(x'|x)$.  The Lagrange multipliers, or equivalently the constraints, provides a freedom to specify the sub-quantum dynamics.  The family of possible sub-quantum dynamics which reproduce the Schr\"{o}dinger equation is potentially infinite, however experiments may constrain these theories once a proper understanding of quantum gravity is achieved.\\

\section{Entropic Dynamics}
In any application of entropic inference, we must supply three pieces of information.  The first of these is the subject matter, the microstates, which is discussed in the next section.  We will then have to supply a prior and any relevant constraints for the problem.
\subsection{The Microstates}
In general treatments of (ED) we consider the positions, $x \in \mathbf{X}$, of $N$ particles in configuration space, $\mathbf{X} = \mathbf{X}_1\times\dots\times\mathbf{X}_N$ which are definite yet \textit{unknown}.  Their unknown values are quantified by a probability density $\rho(x)$.
We also make another assumption, that the particles follow continuous trajectories; the particles move in short steps \cite{Caticha_new}.  The inference framework allows us to find a large change by iterating over many small steps, and thus we only need to find the transition probability for a short step.  The principle of maximum entropy tells us that such a probability should maximize the relative entropy,
\begin{equation}
S[P,Q] = -\int dx'\, P(x'|x)\log\frac{P(x'|x)}{Q(x'|x)}\label{S}
\end{equation}
subject to constraints, however first we must specify the prior $Q(x'|x)$.
\subsection{The Prior}
To incorporate our ignorance about the motion of the particles, we can choose a prior that includes the symmetries in the problem.  Such a prior is a Gaussian,
\begin{equation}
Q(x'|x) \propto \exp\left[-\frac{1}{2}\sum_n\alpha_n\delta_{ab}\Delta x^a_n\Delta x^b_n\right]
\end{equation}
where $\alpha_n$ is some particle dependent constant for which we can take the limit $\alpha_n\rightarrow\infty$ to impose short steps.  Such a prior quantifies the rotational symmetries present in the problem.  In order to break the symmetry, we impose a family of constraints.

\subsection{The Constraints}
Depending on the subject matter, we impose a family of different constraints that incorporate the information that is relevant to the problem.  There are two main classes of problems that we will discuss here, although such a list is not exhaustive.  The first concerns \textit{scalar particles}, or particles without spin, while the second concerns particles \textit{with} spin, and hence the second kind requires additional constraints.  Special cases of either approach concern the study of a single particle \cite{Caticha_Carrara_Spin}, which we will mainly focus on in this paper.  The microstates for a single particle is simply three dimensional space, $\mathbf{X} \subset \mathbb{R}^3$.
\paragraph{The Local $U(1)$ Constraint-}
\indent  The constraints for particles, whether of the scalar or spin variety, also incorporate symmetry information (much like the prior).  We will find that the main symmetry group for scalar particles is $U_{\infty}(1)$, the unitary group in infinite dimensions.  This corresponds to a local gauge symmetry at each position $x \in \mathbf{X}$ and is represented by the following constraint,
\begin{equation}
\langle \Delta x^a\rangle \left[\partial_a\phi(x) - \beta A_a(x)\right] = \kappa(x)\label{constraint_gauge}
\end{equation}
where $\phi(x)$ is a field that has the topological properties of an angle and $\vec{A}$ is a connection field that sets the zero of $\phi(x)$ at each $x$.  The factor $\beta$ is identified with electric charge \cite{Carrara_Caticha_Phases}.    
\paragraph{The $SU(2)$ Constraint - }In order to capture the appropriate rotational properties of the system, we incorporate an additional set of constraints on the motion of the particle\footnote{The results of this section are from joint work with A. Caticha that will appear in \cite{Caticha_Carrara_Spin}.}.  A useful representation of rotations in $\mathbb{R}^3$ is a frame field $\vec{s}_k(x)$ at each point in space, the dynamics of which will be coupled to the particle motion\footnote{The use of frame fields for describing spin has been used throughout the literature \cite{Takabayasi_1, Takabayasi_2, Hestenes_1, Hestenes_2}.}.  Just like the fields $\phi(x)$ and $\vec{A}(x)$, the field $\vec{s}_k(x)$ is entirely epistemic; it is merely a convenient representation of our information about the motion of the particle, there is no assumption that the field $\vec{s}_k(x)$ is ``real.''\\
\indent  A frame $\vec{s}_k(x)$ at a point $x \in \mathbf{X}$ is a triad, $\vec{s}_k(x) = \{\vec{s}_1(x),\vec{s}_2(x),\vec{s}_3(x)\}$, whose individual components span $\mathbb{R}^3$.
Each frame at $x \in \mathbf{X}$ can be constructed by rotating the lab frame, which we denote with the basis vectors $\vec{e}_k = \{\vec{e}_1,\vec{e}_2,\vec{e}_3\}$, through three Euler angles $\{\chi(x),\theta(x),\varphi(x)\}$ which depend on position.  This is performed through the action of a rotor $U(\chi,\theta,\varphi)$,
\begin{equation}
\vec{s}_k(x) = U(x)\vec{e}_kU^{\dagger}(x) = U(\chi,\theta,\varphi)\vec{e}_kU^{\dagger}(\chi,\theta,\varphi)
\end{equation}
where
\begin{equation}
U(\chi,\theta,\varphi) = U_z(\varphi)U_y(\theta)U_z(\chi) = e^{-i\vec{e}_3\varphi/2}e^{-i\vec{e}_2\theta/2}e^{-i\vec{e}_3\chi/2}\label{rotor}
\end{equation}
The frame is said to be oriented along the $\vec{s}_3$ direction with constant magnitude; i.e. $\vec{s}(x) = |\vec{s}| \vec{s}_3(x)$ and $|\vec{s}(x)| = |\vec{s}| = \mathrm{const.}$.  Since the frames can take arbitrary orientation at each $x$, we would like to know how the frame changes its direction from $x$ to $x'$. In the same way that the constraint (\ref{constraint_gauge}) involves the displacement being directed along the gradient of an angle, we incorporate the spin by also coupling the displacement to the gradient of an angle $\zeta_3(x)$
\begin{equation}
\langle \Delta x^a \rangle\partial_a\zeta_3 = \langle\Delta x^a\rangle(\vec{\omega}_a\cdot\vec{s}_3) = \langle \Delta x^a \rangle(\partial_a\chi + \cos\theta\partial_a\varphi) = \kappa'(x)\label{constraint_SU2}
\end{equation}
which is a combination of gradients along the polar angle $\chi(x)$ and the precession angle $\varphi(x)$\footnote{While the derivatives of the angles $\zeta_k(x)$ are well defined, their solutions are in general not integrable.} given in the frame velocity $\vec{\omega}_a$.  Since the motion of the particle is being directed along the $\vec{s}_3$ direction, there is an arbitrariness in the setting of the zero angle of the $\chi(x)$.  This suggests that the $\chi(x)$ in the spin constraint (\ref{constraint_SU2}) plays the same role of a gauge field as the constraint for scalar particles (\ref{constraint_gauge}), and we will see that it is only their joint dynamics that contributes to the evolution of the system.  Thus in cases of a single particle with spin, the constraints (\ref{constraint_gauge}) and (\ref{constraint_SU2}) can be combined into a single constraint which is gauge invariant.
\subsection{The Transition Probability}
Maximizing the relative entropy (\ref{S}) subject to the constraints (\ref{constraint_gauge}) and (\ref{constraint_SU2}) leads to the transition probability
\begin{equation}
P(x'|x) \propto \exp\left[-\frac{\alpha}{2}\delta_{ab}\Delta x^a\Delta x^b + \left(\alpha'(\partial_a\phi - \beta A_a) + \gamma(\vec{\omega}_a\cdot\vec{s}_3)\right)\Delta x^a\right]
\end{equation}
with Lagrange multipliers $\alpha'$ and $\gamma$.  This distribution is Gaussian, and a generic displacement $\Delta x^a$ can be written
\begin{equation}
\Delta x^a = \langle \Delta x^a\rangle + \Delta \omega^a\label{displace}
\end{equation}
where the expected displacement $\langle \Delta x^a \rangle$ is given by
\begin{equation}
\langle \Delta x^a \rangle = \frac{1}{\alpha}\delta^{ab}\left(\alpha'(\partial_b\phi - \beta A_b) + \gamma(\vec{\omega}_b\cdot \vec{s}_3)\right)
\end{equation}
and the fluctuations obey
\begin{equation}
\langle \Delta w^a \rangle = 0 \quad \mathrm{and} \quad \langle \Delta w^a\Delta w^b\rangle = \frac{1}{\alpha}\delta^{ab}\label{fluctuations}
\end{equation}
The Lagrange multiplier $\alpha'$ plays the role of controlling the relative strength of the fluctuations \cite{Bartolomeo_Caticha}.  In the theory of spin the value of $\gamma = 1/2$, while $\beta = e/c$ is proportional to the electric charge\footnote{The (ED) framework offers a unique argument for the quantization of electric charge which is a consequence of the circulation conditions of the spin frame $\vec{s}(x)$ and the single-valuedness of the wave function \cite{Caticha_Carrara,Caticha_Carrara_Spin}.}.  An important quantity is the ratio of the Lagrange multipliers,
\begin{equation}
\frac{\alpha'}{\alpha} = \frac{\gamma}{\alpha} = \frac{\hbar}{m}\Delta t
\end{equation}
The form of $\alpha'$ determines a \textit{class} of motions,
\begin{equation}
\alpha' = \frac{1}{\eta(\Delta t)^n},\quad \gamma = \frac{1}{\xi(\Delta t)^n} \qquad \mathrm{and}\qquad \langle \delta_{ab}\Delta w^a\Delta w^b\rangle = \frac{\hbar}{m}\eta(\Delta t)^{n+1}, \qquad |\Delta w| \propto \sqrt{\frac{\hbar}{m}\eta}(\Delta t)^{(n+1)/2}
\end{equation}
for some integer $n$.  For $n=0$, the particles follow Brownian trajectories, which in the limit of $\eta \rightarrow 0$ and $\xi \rightarrow 0$ recovers the smooth Bohmian trajectories \cite{Bartolomeo_Caticha}.  

\subsection{Entropic Time}
At this point Entropic Dynamics describes a theory of particles which undergo a particular class of motion depending on the choice of constraints (\ref{constraint_gauge}), (\ref{constraint_SU2}).  The next step is to define an \textit{entropic time} \cite{Caticha_EntropicTime} by associating to the equation,
\begin{equation}
\rho(x') = \int dx\, P(x'|x) \rho(x)\label{update}
\end{equation}
a notion of \textit{duration}, supplied by the fluctuations $(\Delta t)$.  The distribution $\rho(x)$ \textit{becomes} the distribution $\rho(x')$, and the procedure in (\ref{update}) has an implicit direction as demonstrated by Bayes' rule\footnote{The update provided by marginalizing over the transition probability $P(x'|x)$ is not necessarily symmetric.  Updating in reverse is constrained by Bayes' rule, $P(x|x') = P(x)P(x'|x)/P(x')$.}.  Much in the way that time is defined in classical mechanics by the free particle -- the free particle moves equal distances in equal time -- entropic time is defined by the free quantum particle; the free quantum particle undergoes equal fluctuations in equal entropic time.\\
\indent  It's often easier to work with the differential form of the integral in eq. (\ref{update}), which can be found to be,
\begin{equation}
\partial_t\rho = -\partial_a(v^a\rho)\label{ent_tra}
\end{equation}
where the velocity $v^a$ depends on the class of motion determined by the constraints and $(\eta,n)$.  For Brownian trajectories, $n=0$, the velocity is,
\begin{equation}
v^a = \frac{\hbar}{m}\delta^{ab}\left(\alpha'(\partial_b\phi - \beta A_b) + \gamma(\vec{\omega}_b\cdot\vec{s}_3) - \partial_b\log\rho^{1/2}\right)
\end{equation}
and equation (\ref{ent_tra}) is the Fokker-Plank equation, which includes the appearance of the osmotic term $\log\rho^{1/2}$.  For the smoother trajectories, $n = 1$, the osmotic term disappears and the velocity becomes $v^a = \langle \Delta x^a \rangle/\Delta t$.  The eq. (\ref{ent_tra}) is a diffusion equation which can be rewritten as a functional derivative, $\partial_t\rho = \delta \tilde{H}/\delta\Phi$ for some functional $\tilde{H}$.
\subsection{Information and Symplectic Geometry}
In order to touch base with quantum mechanics, we investigate the geometric and symplectic structure of the statistical manifold of $\rho(x) \in \mathbf{\Delta}$.  We will not go into detailed discussion here, except to mention a few key points.  The central object of interest is the cotangent bundle (phase space) of $\mathbf{\Delta}$, denoted $T^*\mathbf{\Delta}$.  Once one establishes a symplectic structure $\Omega \in T^*\mathbf{\Delta}$, one can determine the flows in the statistical manifold which preserve it; these are Hamiltonian flows whose generators belong to the symplectic group $Sp_{\infty}(2n)$.  If one also imposes that these flows preserve the metric,  that such flows define isometries, then they are Hamilton-Killing flows whose generators belong to the group $Sp_{\infty}(2n)\cap O_{\infty}(2n)$.  The intersection of these two groups happens to be equal to, $Sp_{\infty}(2n)\cap O_{\infty}(2n) = U_{\infty}(n)$, which is the unitary group, of which the gauge and rotational symmetries provided by the constraints (\ref{constraint_gauge}) and (\ref{constraint_SU2}) are a subset.\\
\indent  The conjugate momenta to $\rho(x)$ ends up being the phase \cite{Carrara_Caticha_Phases}, which for the Brownian case is $\Phi(x)/\hbar = \gamma\chi + \alpha'\phi - \log\rho^{1/2}$.  One can find a Hamiltonian $\tilde{H}$ which reproduces the entropic equation (\ref{ent_tra}),
\begin{equation}
\tilde{H}[\rho,\Phi,\rho_s,\Phi_s] = \int dx\, \left[\frac{1}{2m}\rho \left(\partial_a\Phi + \frac{\rho_s}{\rho}\partial_a\Phi_s -\beta A_a\right)^2  +  \frac{\hbar^2}{8m}\rho\left(\frac{\partial^a\partial_a\rho}{\rho^2} + (\partial_a\vec{s}_3)^2\right) + \rho V + \frac{\beta}{m}\rho \vec{s}\cdot\vec{B} \right]
\end{equation}{
where $\rho_s = \rho\cos\theta$ and $\Phi_s/\hbar = \gamma\varphi$ are conjugate variables incorporating the extra spin degrees of freedom.  Combining the conjugate Hamilton equations for the phases $\Phi,\Phi_s$ we get,
\begin{align}
\partial_t\Phi + \frac{\rho_s}{\rho}\partial_t\Phi_s = -\frac{\delta \tilde{H}}{\delta \rho} - \frac{\rho_s}{\rho}\frac{\delta \tilde{H}}{\delta \Phi_s} = -\frac{1}{2m}&\left(\partial_a\Phi + \frac{\rho_s}{\rho}\partial_a\Phi_s -\beta A_a\right)^2\nonumber\\
 &+ \frac{\hbar^2}{2m}\frac{\partial^a\partial_a\rho^{1/2}}{\rho^{1/2}} - \frac{\hbar^2}{8m}(\partial_a\vec{s}_3)^2 - V - \frac{\beta}{m}\vec{s}\cdot\vec{B}
\end{align}
which is the Hamilton-Jacobi equation and is equal to the local energy $\partial_t\Phi + (\rho_s/\rho)\partial_t\Phi_s = (\hbar/2)\vec{\omega}_t\cdot \vec{s}_3$.  The other Hamilton equation leads to,
\begin{equation}
\mathcal{D}_t\vec{s} = -\frac{\hbar}{2m}\left\{\left(\frac{1}{\rho}\partial^a(\rho\partial_a\vec{s}_3)) + \beta|\vec{s}|\vec{B}\right)\times \vec{s}_3\right\}\cdot \vec{e}_3
\end{equation}
which is the spin precession equation and $\mathcal{D}_t = \partial_t + v^a\partial_a$ is the convective derivative.\\
\indent  The equivalence of the intersection of the symplectic and orthogonal groups to the unitary group also results in another symmetry, the group $GL(2n,\mathbb{C})$, which is the complex linear group whose generators preserve a complex structure $J$.  The existence of a complex structure suggests we can use complex coordinates in the phase space.  Such a set of coordinates which are a canonical transformation are,
\begin{equation}
\psi_{\pm} = \rho_{\pm}^{1/2}e^{i\Phi_{\pm}/\hbar} \qquad \mathrm{and}\qquad i\hbar \psi_{\pm}^{\dagger} = i\hbar\rho_{\pm}^{1/2}e^{-i\Phi_{\pm}/\hbar}
\end{equation}
where $\rho_{\pm} = (1/2)(\rho \pm \rho_s)$ and $\Phi_{\pm} = (\Phi \pm \Phi_s)$.  The Hamiltonian becomes,
\begin{equation}
\tilde{H}[\psi_{\pm},i\hbar\psi^*_{\pm}] = \int dx\, \left(-\frac{\hbar^2}{2m}\psi^*_{\pm}(\partial_a - (i/2)A_a)^2\psi_{\pm} + \psi^*_{\pm}V\psi_{\pm} + \psi_{\pm}^*(B_x \mp iB_y)\psi_{\mp} \pm \psi_{\pm}^*B_z\psi_{\pm}\right)
\end{equation}
and the associated Hamilton's equation,
\begin{equation}
i\hbar\partial_t\psi_{\pm} = \frac{\delta \tilde{H}}{\delta\psi^*_{\pm}} = -\frac{\hbar^2}{2m}(\partial_a - (i/2))^2\psi_{\pm} + V\psi_{\pm} + (B_x \mp iB_y)\psi_{\mp} \pm B_z\psi_{\pm}
\end{equation}
is the Schr\"{o}dinger equation for the $(\pm)$ components of the Pauli equation.  In the limit that the variables $\theta,\varphi$ are not dynamical, the Pauli equation reduces to the Schr\"{o}dinger equation for a scalar particle.

\section{Entropic Trajectories}
Entropic trajectories are a generalization of the trajectories assumed by (BM) and (NSM).  In Bohmian mechanics these trajectories are smooth, with well defined velocities, that are also constrained to never cross.  They are determined from the probability flow, which for scalar particles is given by
\begin{equation}
\frac{d\vec{x}_n}{dt} = \vec{v}_n, \qquad \mathrm{where} \quad \vec{v}_n = i\frac{\hbar}{2m_n}\left(\frac{\vec{\nabla}_n\psi}{\psi} - \frac{\vec{\nabla}_n\psi^*}{\psi^*}\right)\label{bohm}
\end{equation}
where $\vec{x}_n$ is the position of the $n$th-particle.  The Bohmian velocity $\vec{v}_n$ is equivalent to the drift velocity $\vec{b}$ in (ED).  In (NSM) the equation of motion for the particles is given by the stochastic equation,
\begin{equation}
d\vec{x} = b(\vec{x},t)dt + dw(t)\label{nelson}
\end{equation}
The velocity from (\ref{bohm}) is not defined in (\ref{nelson}) in the standard limit calculus sense and hence one can only evaluate finite differences. 
In Entropic Dynamics, it is the displacement (\ref{displace}) which determines the motion of the particles.  The displacement contains the fluctuation term, which is stochastic, hence the limit in (\ref{bohm}) is not always defined.  
While one can evaluate the limit using stochastic calculus, we will relegate that discussion to a future paper.  For the collection of simulations in this paper, we will simply use a unit fluctuation $\Delta \tilde{w}$ in place of the Wiener process $\Delta \bar{w}$, which simulates a random walk on the unit sphere.  A finite time step is simulated by providing a duration $\Delta t$ and some prescribed values of $n$ and $\eta$.  The displacement for Brownian motion is then found from,
\begin{equation}
\Delta x^a = b^a\Delta t + \sqrt{\frac{\hbar}{m}\eta}(\Delta t)^{1/2}\Delta \tilde{w}^a\label{displace_2}
\end{equation}
where $b^a$ is given from the Bohmian limit.  In the examples below, eq. (\ref{displace_2}) is integrated using the standard $4$th-order Runge Kutta method.  
\subsection{The Double-Slit Experiment}
The double slit experiment \cite{Jonsson, Gondran_2} is a special case where the wave function can be solved exactly by assuming that each slit produces a Gaussian wave packet with a width equal to the width of the slit, $\sigma_0$, and that the total wave function is represented by a super-position of each packet,
\begin{equation}
\psi_i(x,y,t) = (2\pi\sigma^2)^{-1/4}\exp\left[-\frac{(y_i - d - \hbar k_yt/m)^2}{4\sigma\sigma_0} + i\left\{\left(k_y(y_i - d) - \frac{\hbar k_y^2 t}{2m}\right) + \left(k_xx - \frac{\hbar k_x^2 t}{2m}\right)\right\}\right]
\end{equation}
where $\hbar k_y = m v_y$, $y_a = y$, $y_b = -y$, and $2d$ is the distance between the slits.  The factor $\sigma$ is $\sigma = \sigma_0\left(1 + \frac{i\hbar t}{2m\sigma_0^2}\right)$.
Each wave function $\psi_i(x,y,t)$, is found from integrating the Schr\"{o}dinger equation for a free particle, $i\hbar\partial_t\psi_i = -(\hbar^2/2m)\vec{\nabla}^2\psi_i$ with an initial Gaussian wave function.  The total wavefunction is the superposition, $\Psi(x,y,t) = N\left[\psi_a(x,y,t) + \psi_b(x,y,t)\right]$.
We simulate the trajectories of electrons, $m = m_e$, with an initial velocity in the $x$ direction of $2\times 10^6 m/s$ and random initial positions along the $y$ direction sampled according to the initial Gaussian distribution with standard deviation equal to the slit width, $\sigma_0 = 10^{-6}m$ and with distance between the slits $d = 5\sigma_0 = 5\times 10^{-6}m$.  The initial velocity in the $y$ direction is set to zero and the distance to the screen is $x_f = 0.2m$.
\begin{figure}[H]
	\centering
	\includegraphics[width=.65\linewidth]{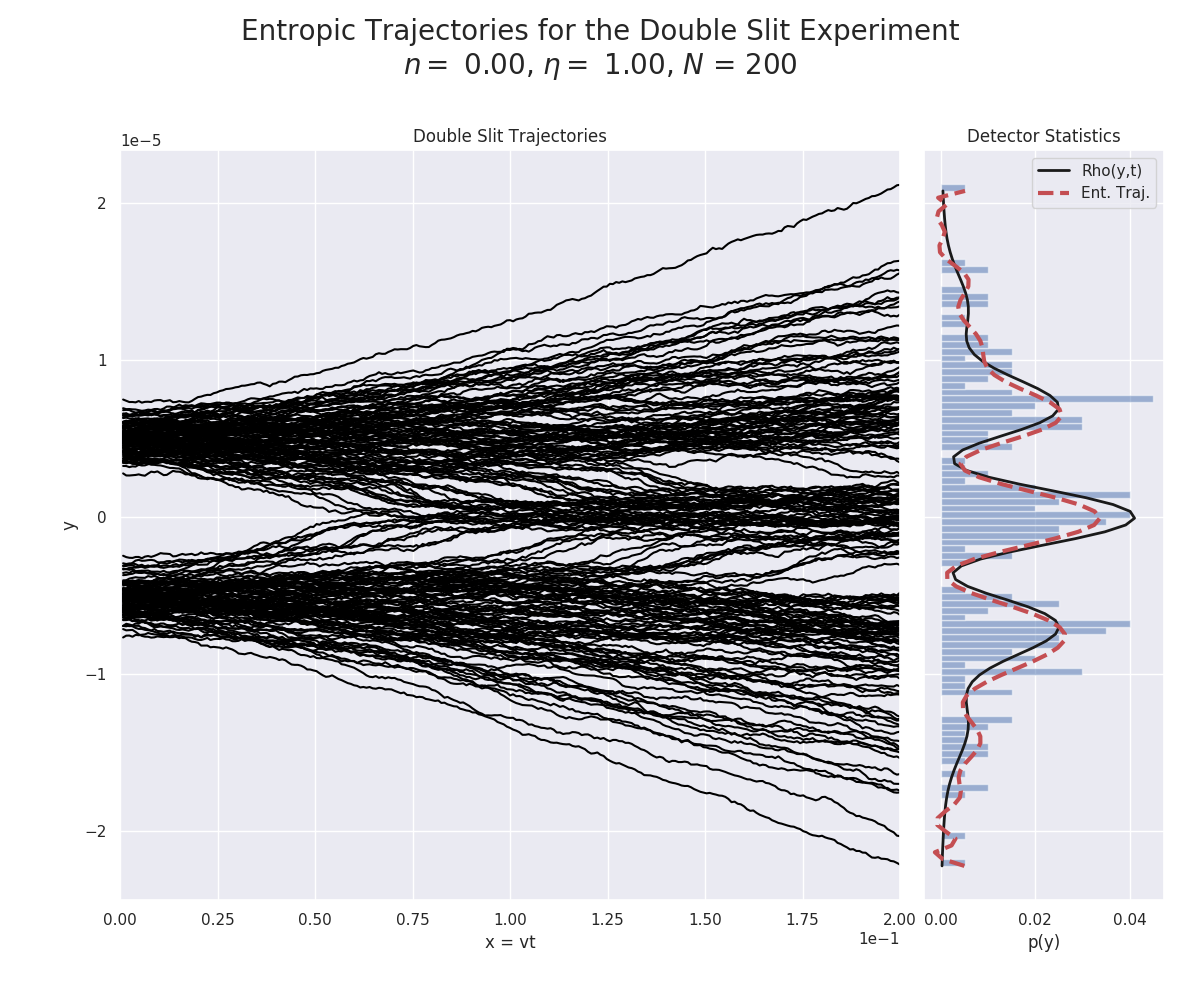}\\
	\includegraphics[width=.65\linewidth]{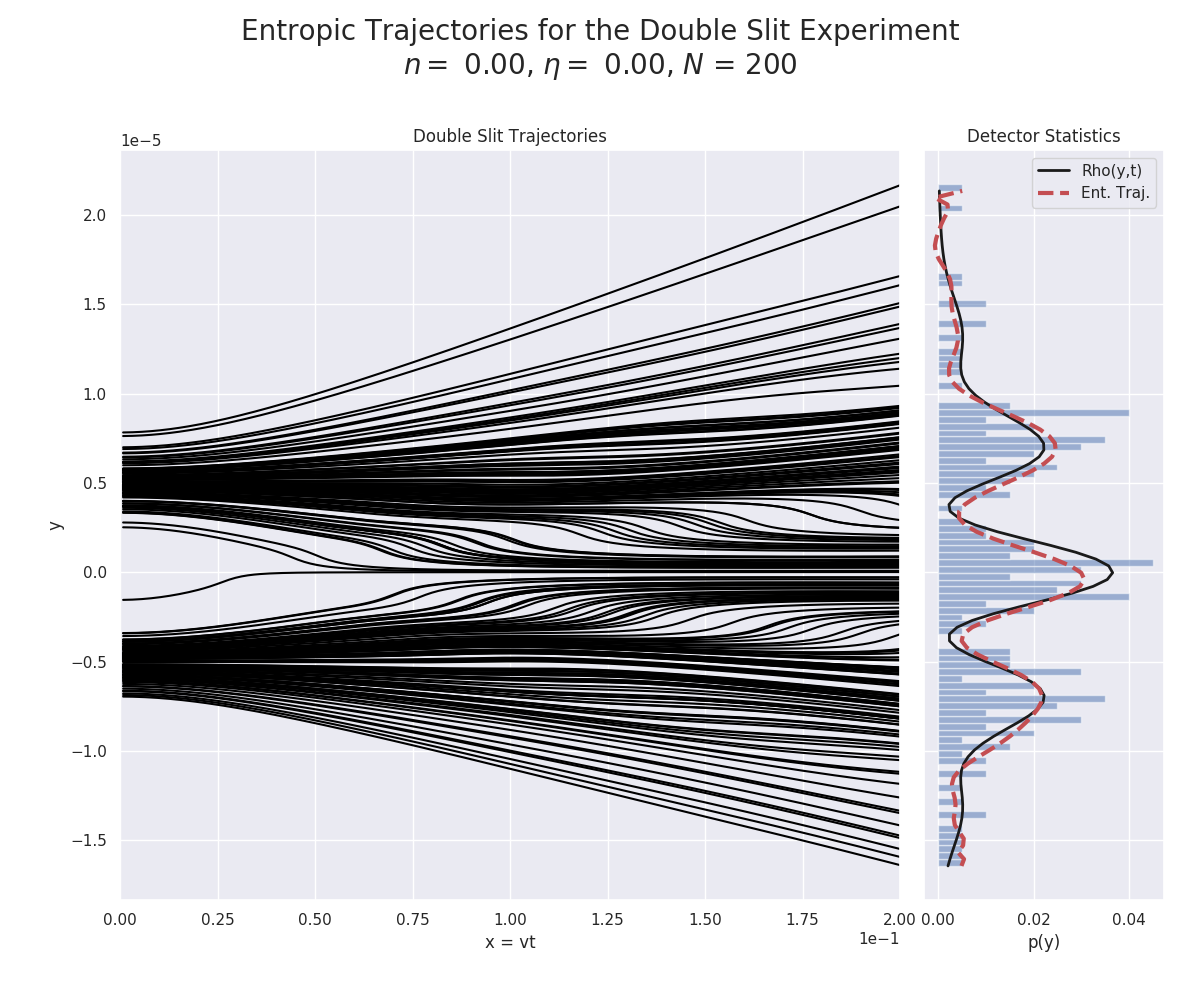}
	\caption{Entorpic trajectories for the double slit experiment with $n=0$ and $\eta = 1,0$ for $N=200$ particles.  The black curve ($Rho(y,t)$) is the probability distribution determined from the wave function at the detector screen, while the red curve (Ent. Traj.) is the interpolated distribution from the detector statistics using a fitting polynomial of order 15.}
\end{figure}
One can see that the value of $\eta = 1$ generates fluctuations which give rise to similar statistics as the Bohmian limit.

\subsection{The Stern-Gerlach Experiment}
In a similar way to the Double-slit experiment, we can solve the Pauli equation in the case of the Stern-Gerlach experiment \cite{SternGerlach} by making a few approximating assumptions \cite{Gondran_3}.  Following the arguments in \cite{DHK_Spin, DST}, we assume that the Stern-Gerlach magnet produces a magnetic field, $\vec{B} =  (B_0 + zB_0')\hat{z}$, within a region $\Delta x$ and is assumed to be zero outside this region.  Given an initial particle velocity $v_x$ along the $x$ direction, the particle remains in the magnetic field for a time $\Delta t = \Delta x/v_x$.  After the particle leaves the magnetic field, the spinor wave function breaks up into two packets which can be solved for all $t$\footnote{For detailed calculations see \cite{Gondran_1,Gondran_2,Gondran_3,Carrara_ET}.} as,
\begin{equation}
\Psi(z,t+\Delta t) = (2\pi \sigma_0)^{-1/2}\begin{pmatrix}
\cos\frac{\theta_0}{2}\exp\left[-\frac{(z - \Delta_z - ut)^2}{4\sigma_0^2} + \frac{i}{\hbar}\left(muz + \hbar \varphi_+\right)\right]\\
\sin\frac{\theta_0}{2}\exp\left[-\frac{(z + \Delta_z + ut)^2}{4\sigma_0^2} - \frac{i}{\hbar}\left(muz - \hbar \varphi_-\right)\right]
\end{pmatrix}
\end{equation}
where $\theta_0$ is the initial azimuthal angle for $\vec{s}_3$ with respect to the $z$ axis, $u$ is the packet velocity in the $z$ direction, $\Delta_z = \mu_BB_0'(\Delta t)^2/2m$ and $\varphi_{\pm} = \pm\varphi_0/2 \mp\mu_BB_0\Delta t/\hbar - \mu_B^2 (B_0')^2(\Delta t)^3/6m\hbar$, where $\mu_B$ is the Bohr magneton.  The width $\sigma_0$ of the initial packet is set to the (SG) device opening of $\sigma_0 = 10^{-4} m$.  We simulate the trajectories of silver atoms with mass $m \approx  1.8e-25$ and an initial velocity along the $x$ direction of $500 m/s$ sampled according to the initial wave packet.  The magnetic field parameters are set to $B_0 = 5 T$ and $B_0' = 10^3 T/m$.  Assuming the particle remains in the magnetic field for a time $\Delta t = 2\times 10^{-5}s$, the factors $\Delta z = 10^{-5}m$ and $u = 1 m/s$.\\
\begin{figure}[H]
 	\centering
 	\includegraphics[width=.65\linewidth]{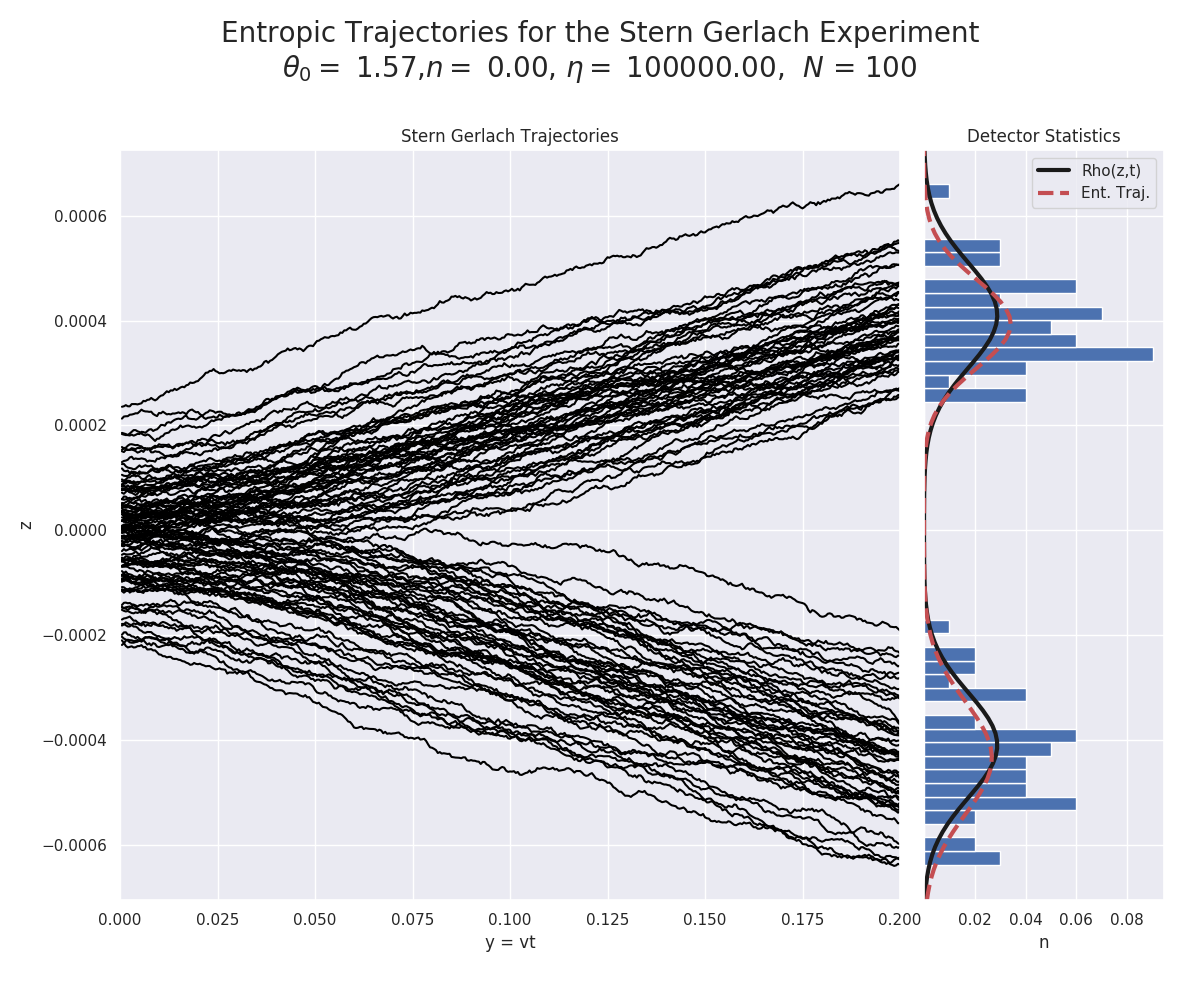}\\
 	\includegraphics[width=.65\linewidth]{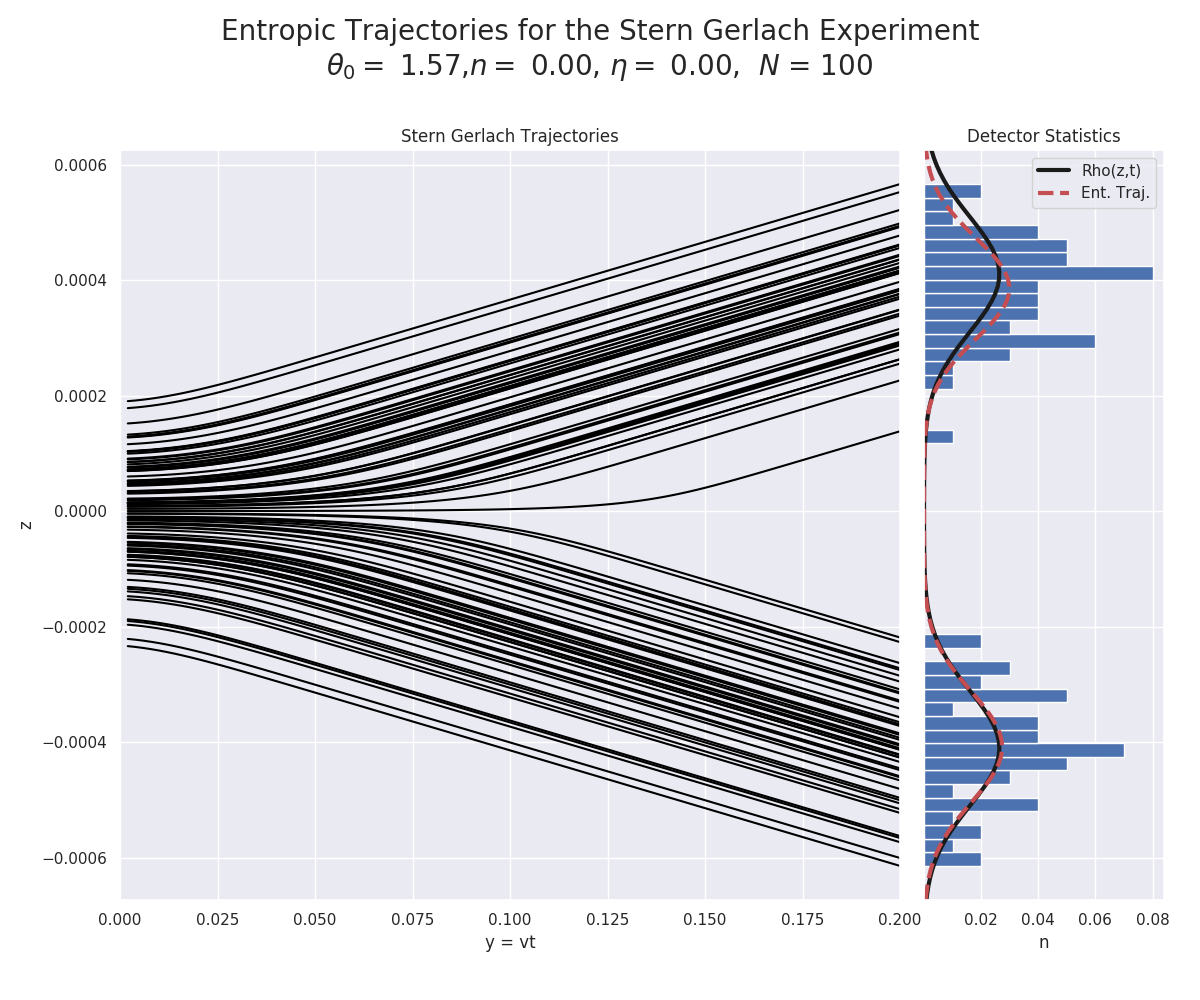}
 	\caption{Entropic trajectories for the Stern-Gerlach experiment with $n=0$ and $\eta = 0, 10^5$ for $N=100$ particles.}
\end{figure}

\begin{figure}[H]
	\centering
	\includegraphics[width=.65\linewidth]{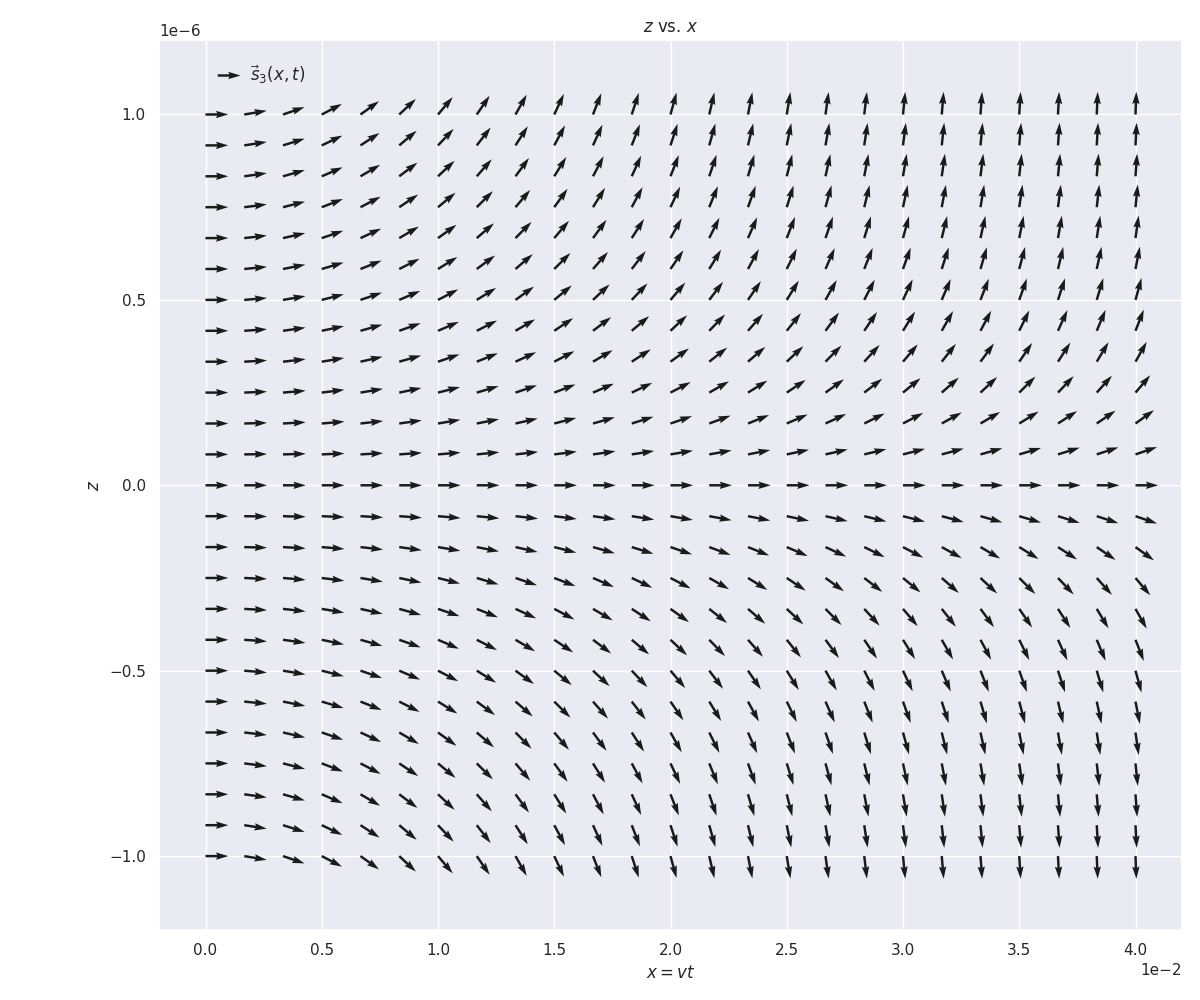}
	\caption{Starting with the initial condition $\theta_0 = \pi/2$, we show the evolution of the direction of the spin frame over $x$ and $t$ with respect to the $xz$-plane.}
\end{figure}
Unlike in the (DS) experiment, the fluctuations are suppressed in this example since the mass of silver is so much larger than electrons, hence the need for $\eta \propto 10^5$ before we start to see Brownian motion.\\
\indent  As we've stated in the introduction and throughout, the spin is entirely epistemic and is not assumed to be a \textit{property} of the particle, but rather a property of its motion.  Much like in \cite{DHK_Spin}, the above example shows how the epistemic spin frame evolves over space and time.  The two-valuedness of spin measurements is not the same type of quantization that is attributed to the particle, but rather just a consequence of measurement, as can be seen from the trajectories. By measuring the particle up or down on the screen, we then assume that the spin must have been up or down at the magnet.  From fig. 3 however, the up and down trajectories are \textit{created} by the (SG) magnet and initially the spin is only up in the $x$ direction.

\section{Discussion}
The (ED) formalism allows for generalized particle trajectories which are not a priori realizable in other foundational approaches.  This freedom is granted by (ED)'s foundation in entropic inference, which requires us to supply information about the symmetries in the problem through constraints.  As we have seen, the Bohmian and Brownian limits are easily attainable, and both give consistent results with respect to experiment.  It still remains an open question as to what classes of sub-quantum dynamics are allowable in quantum mechanics, and ultimately in quantum gravity.  While at the moment we cannot answer the latter, we will address the former question in longer paper which will extrapolate further on the discussion from section three.  

\section*{Acknowledgement}We would like to thank A. Caticha, J. Ernst, S. Ipek, P. Pessoa and K. Vanslette for insightful conversations.

\bibliography{maxentbib.bib}
\end{document}